\documentclass[conference]{IEEEtran}
\usepackage{cite}
\usepackage{amsmath,amssymb,amsfonts}
\usepackage{graphicx}
\usepackage{subfigure}
\usepackage{textcomp}
\usepackage{xcolor}

\graphicspath{ {./content/imgs/} }

\usepackage{hyperref}
\usepackage{algorithm, algorithmicx, algpseudocode}
\usepackage{multirow}

\usepackage{mathtools}
\DeclareMathOperator*{\argmax}{arg\,max}
\DeclareMathOperator*{\argmin}{arg\,min}

\usepackage[english]{babel}

\begin{document}
	
	\title{Fast off-the-grid sparse recovery with over-parametrized projected gradient descent
	}
	
	\author{\IEEEauthorblockN{Pierre-Jean Bénard\textsuperscript{1,*},Yann Traonmilin\textsuperscript{1},Jean-François Aujol\textsuperscript{1} 
	 \\
			\textit{\textsuperscript{1}Univ. Bordeaux, Bordeaux INP, CNRS,  IMB, UMR 5251,F-33400 Talence, France}}
			\textsuperscript{*}contact: pierre-jean.benard@math.u-bordeaux.fr		
	}
	
	\maketitle
	
	\begin{abstract}
		We consider the problem of recovering off-the-grid spikes from Fourier measurements. Successful methods such as sliding Frank-Wolfe and continuous orthogonal matching pursuit (OMP) iteratively add spikes to the solution then perform a costly (when the number of spikes is large) descent on all parameters at each iteration. In 2D, it was shown that performing a projected gradient descent (PGD) from a gridded over-parametrized initialization was faster than continuous orthogonal matching pursuit. In this paper, we propose an off-the-grid over-parametrized initialization of the PGD based on OMP that permits to fully avoid grids and gives faster results in 3D.
	\end{abstract}

	\begin{IEEEkeywords}
		spike super-resolution, non-convex optimization, over-parametrization, projected gradient descent, continuous orthogonal matching pursuit
	\end{IEEEkeywords}
	
	\section{Introduction}

	Let $ x_{0} $ be an off the grid sparse signal over $ \mathbb{R}^{d}$. Such signals can be modeled as a sum of $k$ Dirac measures:
	\begin{equation}\label{eq:initial_signal}
		x_{0} = \sum_{i = 1}^{k} a_{i} \delta_{t_{i}}
	\end{equation}
	where $ a  =(a_1,\ldots,a_k)\in \mathbb{R}^{k} $ are the amplitudes and $t= (t_1,\ldots,t_k) \in \mathbb{R}^{k \times d} $ are the locations of the spikes.  We observe this signal through $ m $ Fourier measurements at frequencies $ \omega= (\omega_1,\dots,\omega_m) \in \mathbb{R}^{d \times m} $. We write this as $y = A x_{0}$
	with $ A $ the corresponding linear operator from the space $\mathcal{M}$ of finite signed measures over $\mathbb{R}^{d}$ to $\mathbb{C}^{m} $. Note that we consider the noiseless case for the sake of clarity.
	
	A way to recover the true signal $ x_{0} $ is to find the minimizer of a non-convex  least-squares problem:
	\begin{equation}\label{eq:least_square}
		x^* \in \argmin_{x \in \Sigma_{k,\epsilon}} \| A x - y \|_{2}^{2}
	\end{equation}
	where $ \Sigma_{k,\epsilon} $ is a set modeling a separation constraint between spikes. Theoretical guarantees for the recovery of $x_0$ with \eqref{eq:least_square} have  been given by Gribonval and al. in \cite{gribonval2021compressive}, e.g. when frequencies are drawn with a well chosen Gaussian distribution and $m\gtrsim O(k^2d polylog(k,d))$, we have that $x^*=x_0$. Generalizing in the spectral line estimation \cite{Bourguignon07}, in the case of a continuous observation of the frequencies up to a cut-off value, it has been shown in \cite{candes2014towards} that we can recover $ x_{0} $ if it obeys to a separation constraint. Practically, continuous  orthogonal matching pursuit \cite{keriven2017compressive} (OMP) has been successful at minimizing~\eqref{eq:least_square}. Continuous OMP is inspired by on the grid  OMP \cite{pati1993orthogonal} (derived from the Matching Pursuit algorithm \cite{mallat1993matching}) for which theoretical success guarantees have been described in \cite{Tropp:greed_is_good,tropp2007signal}. Success in the continuous case have been shown in cases  that do not fit all practical applications (such as Dirac recovery from random Gaussian Fourier measurements) \cite{elvira2019omp,elvira2019does}. Another way to estimate $x_0$ is to add a regularization term (total variation of measures) to \eqref{eq:least_square}.
	The regularized functional can then be solved practically with the sliding Frank-Wolfe algorithm \cite{denoyelle:hal-01921604:sliding-frank-wolfe}. 
	
	Both Continuous OMP and Sliding Frank-Wolfe iteratively add a spike then perform a  descent on all parameters (amplitudes and positions). This descent step that we  call the sliding step in both cases is where most of the calculations are made and it makes this class of method computationally heavy when the number of spikes is large.

	The geometry of minimization~\eqref{eq:least_square} in the parameter space has been studied in \cite{traonmilin:hal-01938239}. Consider a set of parameters $\Theta_{k, \epsilon} $ where
	\begin{equation}\label{key}
		\begin{aligned}
			\Theta_{k, \epsilon} {}:={} & \left\{ \theta = (a_{1}, \dots, a_{k}, t_{1}, \dots, t_{k}) \in \mathbb{R}^{k(d + 1)} \right. , \\
			& \: \forall i, j \in \{ 1, \dots, k \}, i \neq j, \| t_{i} - t_{j} \|_{2} > \epsilon \big\}.
		\end{aligned}
	\end{equation}
	We can rewrite \eqref{eq:initial_signal} with the variable $\theta$:
	\begin{equation}
		x_0 = \phi(\theta) := \sum_{i=1}^{k} a_i \delta_{t_i} \in \Sigma_{k, \epsilon}.
	\end{equation}
	The unknown $ x_{0} $ belongs to the low dimensional model 
	\begin{equation}\label{separation_space}
		\Sigma_{k, \epsilon} {}:={} \left\{ \sum_{i = 1}^{k} a_{i} \delta_{t_{i}}, (a_{i}, t_{i}) = \theta \in \Theta_{k, \epsilon} \right\}.
	\end{equation}
	\setlength{\arraycolsep}{5pt}
	Problem~\eqref{eq:least_square} can then be equivalently written as
	\begin{equation}
		\theta^* \in \argmin_{\theta \in \Theta_{k, \epsilon}} g(\theta) \quad \text{ with } g(\theta) := \| A \phi(\theta) - y \|_2^2.
	\end{equation}

	It has been shown that a simple gradient descent converges to a global minimum as long as it is initialized within a basin of attraction of this minimum that has an explicit size. It has been shown that this size increases with respect to the number of measurements. In \cite{traonmilin:projected_gradient_descent}, it was shown that a single projected gradient descent initialized by an over-parametrized back-projection of the measurements on a grid  permits the recovery of a large number of spikes in 2D with improved calculation times compared to Sliding Continuous OMP. However, the use of a grid is an obstacle to the generalization of this method to domains of higher dimensions.

	\noindent \textbf{Contributions.} In this paper, our main contribution is a fast fully grid-less method for recovering sparse signals on domains of any dimension. We propose to use an over-parametrized greedy off-the-grid initialization based on Continuous OMP without a sliding step. Then we perform a projected gradient descent that provides the final estimate. We provide experiments in 2D and 3D that show the success of the algorithm and up to an improvement of five times in calculation times compared to Sliding Continuous OMP. It must also be noted that with recent advances on the study of gradient descent in this context, a full proof of convergence of the proposed algorithm seems now accessible.  
	
	\section{Over-parametrized Continuous OMP and Projected Gradient Descent}
	In this section, we describe our method which consists in two parts: a projected gradient descent and a greedy over-parametrized initialization off the grid.
 For each iteration of the projected Gradient Descent, it performs a projection to keep $ \theta $ in the space $ \Theta_{\epsilon} $. The Gradient Descent operates on all parameters (i.e. $ \theta $) of the estimated signal.
	
	\subsection{Projected gradient descent}
	The projected gradient descent (PGD), as introduced in \cite{chen2015fast}, or in spectral compressed sensing in \cite{cai2018spectral} and for this specific case in \cite{traonmilin:projected_gradient_descent}, is a way to perform a simple ``descent'' algorithm on the function $g$ while guaranteeing $ \theta \in \Theta_{k,\epsilon}$, i.e. our estimate of $x_0$ is in $\Sigma_{k,\epsilon}$.  
	
	Given an initialization $\theta_{init} \in \mathbb{R}^{k_{init}(d+1)}$, we iterate
	\begin{equation}\label{eq:pGD}
		\theta^{(n+1)} = P_{\Theta_{\epsilon}} (\theta^{(n)} - \tau_{n} \nabla g(\theta^{(n)}))
	\end{equation}
	with $ \tau_{n} $ the step size at the $n$th iteration and $ P_{\Theta_{k_{n}, \epsilon}} $ is the projection on the separation constraint for a signal made of $k_n$ spikes at gradient step $n$.
	Practically, a heuristic is used to perform the projection $ P_{\Theta_{k_{n}, \epsilon}} $ in \cite{traonmilin:projected_gradient_descent}: if two spikes are too close i.e. their distance from one another is below a threshold, they are merged to form one spike. To merge them, we add the amplitudes and we take the barycenter of the locations weighted by the amplitudes. 
	
	Of course, the critical step for the use of only one pass of this algorithm is that the over-parametrized initialization is close enough to $\theta^*$. Practically, in 2D it is possible to use a simple hard thresholding of the back-projection of measurements on a grid. Given a grid $\Gamma$, we can calculate the back-projection $z_{\Gamma}\in \mathcal{M}$ of $y$ on the grid $\Gamma$ as  
	\begin{equation}\label{eq:backprojection}
		z_{\Gamma} =\sum_{s_{j} \in \Gamma} z_{\Gamma, j} \delta_{s_{i}} \quad \text{with} \; z_{\Gamma, j} = \sum_{l=1}^m y_{l}  e^{j \langle \omega_{l}, s_{j} \rangle}.
	\end{equation}
	It is then possible to extract greedily an initial $\theta_{init}$ from the largest amplitudes in $z_{\Gamma}$~\cite{traonmilin:projected_gradient_descent}. While successful in 2D for the initialization of PGD (and useful for an easy visualization of the quality of the sampling, see next section), the curse of dimension limits the possibility to extend such gridded method in domains of higher dimensions.
	
	\subsection{Initialization with Over-parametrized Continuous}
	The over-parametrized COMP (Continuous Orthogonal Matching Pursuit) is an alteration of  Sliding COMP. As described in Algorithm~\ref{alg:SCOMP}, it iterates over the number of total spikes to add. It finds a location where the spike maximizes a correlation with the residue. Then it performs an update of the amplitudes and it refreshes the residue. The idea is to skip the sliding part to yield an initialization close enough to the true solution for the PGD. 
	If the norm of the residue decreases below a threshold or if the number of spikes to add passes a certain value, then we stop the iterations.
	
	The role of  over-parametrization is to compensate the inaccuracies induced at each iteration by the removal of the sliding step. This can be viewed as an off-the-grid generalization of the previously proposed gridded greedy initialization \cite{traonmilin:projected_gradient_descent}.  The over-parametrized COMP is described in Algorithm~\ref{alg:SCOMP}.
	\begin{algorithm}[H]
		\caption{Continuous Orthogonal Matching Pursuit algorithm. The Sliding COMP performs state of the art spike recovery. We use an over-parametrized  COMP without sliding as initialization of our PGD.}
		\label{alg:SCOMP}
		
		\begin{algorithmic}
			\Procedure{COMP}{$A, y, K, \mathtt{is\_sliding}$}
			
			\State $ r^{(0)} \gets y $
			\State $ t^{(0)} \gets \{\} $
			
			\For{$k = 1 \to K$}
				\State $ T \gets \argmax_{t} \langle A \delta_{t}, r^{(k - 1)} \rangle$
				\State $ t^{(k)} \gets t^{(k-1)} \cup \{ T \}$
				\State $ a^{(k)} \gets \argmin_{a} \| A \sum_{i=1}^{k} a_{i} \delta_{t_{i}^{(k)}} - y \|_{2}^{2} $ \Comment{Update of the amplitudes}
				
				\If{$ \mathtt{is\_sliding} $}
				\State $ a^{(k)}, t^{(k)} \gets \texttt{descent}(g, a^{(k)}, t^{(k)})$ 
				\EndIf
				
				\State $ r^{(k)} \gets y - A \sum_{i=1}^{k} a_{i}^{(k)} \delta_{t_{i}^{(k)}} $ \Comment{Update of the residue} \label{alg:residue_def}

			\EndFor
			
			\State \textbf{return} $ a^{(K)}, t^{(K)} $
			
			\EndProcedure
		\end{algorithmic}
		
	\end{algorithm}
	
	To chose the amount of over-parametrization $k_{init}$, it was proposed  to chose a fixed multiple of an estimated true number of spikes \cite{traonmilin:projected_gradient_descent}.
	In our new method, we add spikes until the residue reaches a threshold or when it stops decreasing. This method makes sure that adding more spikes with our chosen COMP would not bring more information in the initialization of the projected gradient descent (as the residue would not decrease further). 
	It must be noted that the more robust Sliding COMP with Replacement was proposed in \cite{keriven:hal-01165984}. In the version with replacement, $2k$ spikes are estimated to produce an estimate with $k$ spikes (making the last sliding steps more costly). We choose to compare ourselves with the faster version without replacement as the main objective of our algorithm is to provide a fast estimation (i.e. we place ourselves in the least favorable case for execution times comparison).
	
	\subsection{Complexity}
	
	To estimate the time gained with our method over Sliding COMP, we  analyze qualitatively the  number of iterations in the descent step  (which is the most time consuming) in each method. For the Sliding COMP method, the execution time  is 
	\begin{equation}\label{eq:sliding_COMP_steps}
		\mathtt{T}_{SCOMP}= \mathcal{O}( N \times (1 +  \dots + K)) = \mathcal{O}(N K^{2})
	\end{equation}
 	with $ N $ the number of iteration in the gradient descent and $ K $ the number of spikes. For over-parametrized COMP, the execution time is
	\vspace{-5pt}
 	\begin{equation}\label{eq:PGD_steps}
 			\mathtt{T}_{PGD}  = \mathcal{O}(K \sum_i N_{i} C_i )
 	\end{equation}
	with $ N_{0} = \sum_i N_i $ the number of iteration in the Projected Gradient Descent method, $N_i$ the number of iterations between each projection and $ C_i $ the constant symbolizing the over-parameterization between each projection, i.e. after the $i$-th projection, there are $K_i=C_i K$ spikes left.
	
	As the over-parametrization factors $C_i$ are generally low compared to $K$, we observe that when the number of spikes $K$ grows, the PGD becomes more interesting than Sliding COMP. We also remark  PGD is faster if projections happen early in the descent. 
	Moreover, when compared to the grid-based initialization from \cite{traonmilin:projected_gradient_descent}, the over-parametrized COMP  complexity is no longer  exponential in the dimension of the support of the spikes.
	
	\section{Experiments}
	
	In this section, we compare our over-parametrized COMP + PGD with  Sliding COMP. We study these methods in 2D and 3D in a compressed acquisition example. For both 2D and 3D cases, the signal is composed of a hundred of spikes. The domain in which the locations of the spikes are is either $ [0, 1]^{2} $ for the 2D case or $ [0, 1]^{3} $ for the 3D case. 
	The code for this experiments is available for download at \cite{benard2022code}.	
	
	\subsection{Comparisons on the 2D case}
	For this case, the signal has the following properties: $ k = 100 $, the minimum distance between two spikes is at least $ \epsilon_{\text{dist}} = 0.015 $. Moreover, the amplitudes follow a uniform distribution $ U([1, 5]) $. In addition, the number of measurements taken is $ m = 40 \times k = 4000 $. The frequencies of measurements follow a normal distribution $ \mathcal{N}(0, c^{2}) $, with $ c = \frac{1}{0.02} \approx \frac{1}{\epsilon_{\text{dist}}} $. The choice of the variance is primordial for the success of recovery. Indeed, The frequencies at which the signal is observed can be too high or too low resulting in a bad approximation of the signal as described in \cite{chatalic2020learning}. In practice, an estimation of $ \epsilon_{\text{dist}} $ is needed to set $ c $.
	
	\begin{figure}[htbp]
		\centering
		\subfigure[]{
			\includegraphics[width=0.45\columnwidth]{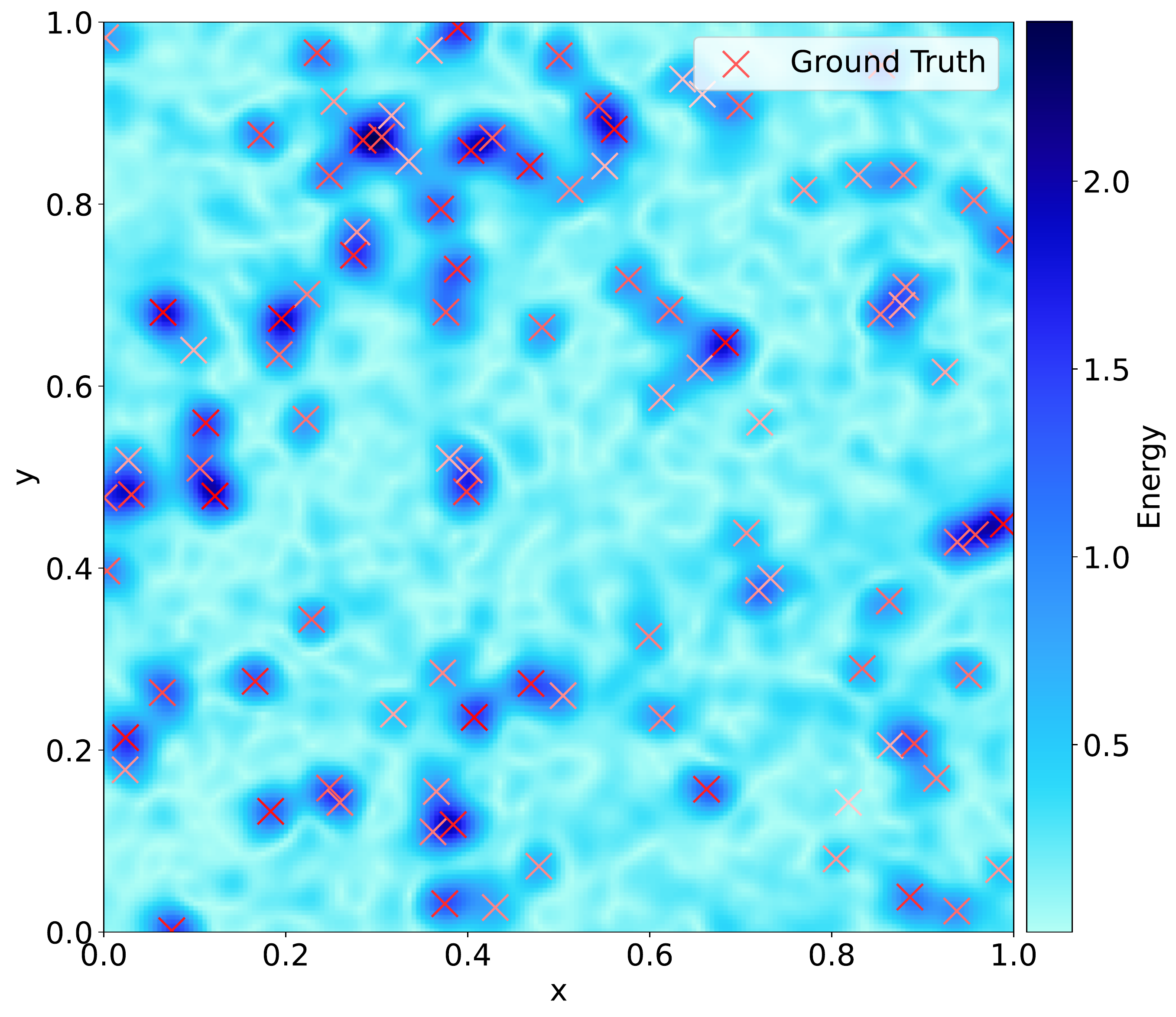}
			\label{fig:2D_gd_energy}
		} 
		\subfigure[]{
			\includegraphics[width=0.43\columnwidth]{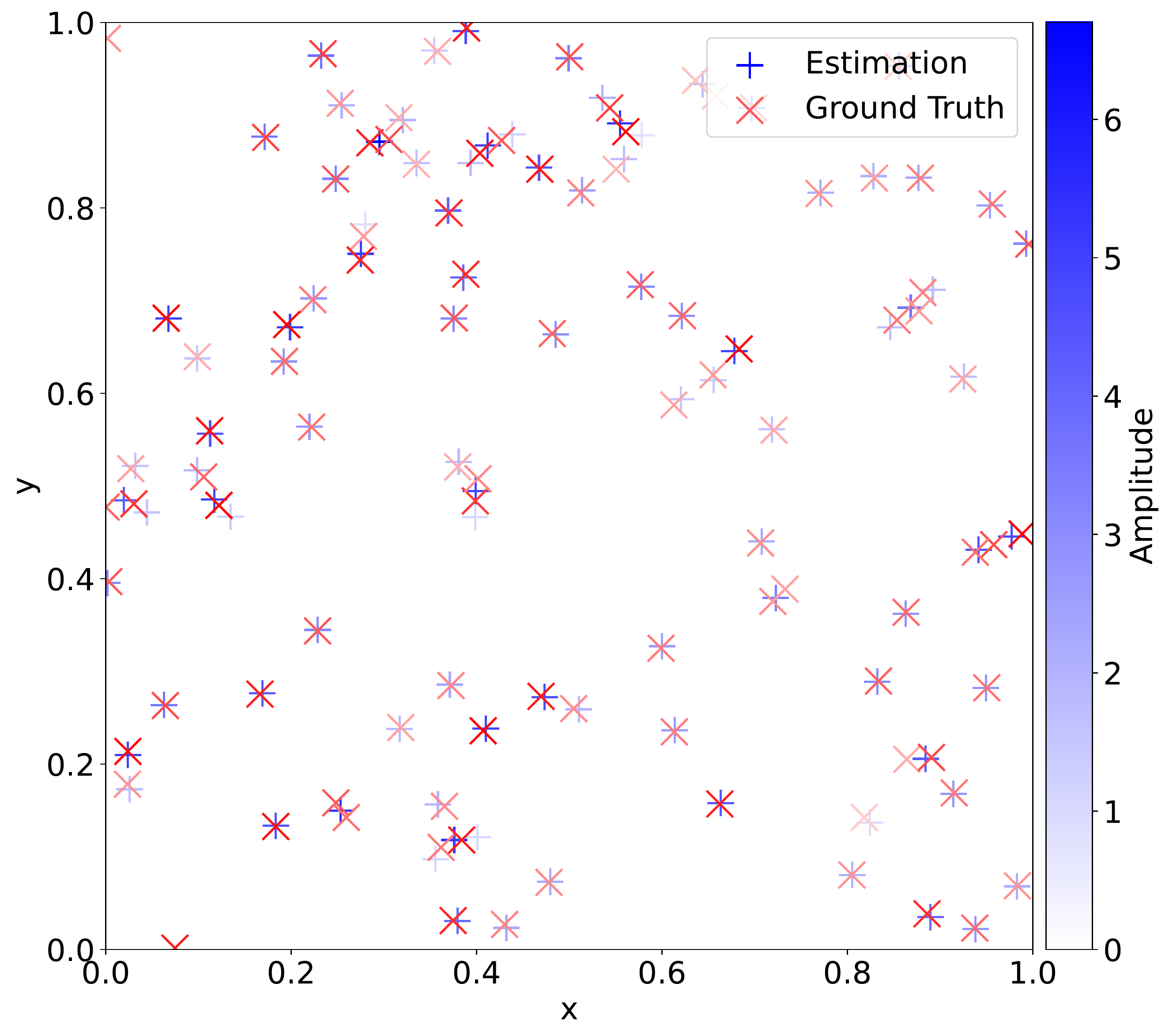}
			\label{fig:2D_COMP_gd_esti}
		} \\[-2ex]
		\subfigure[]{
			\includegraphics[width=0.45\columnwidth]{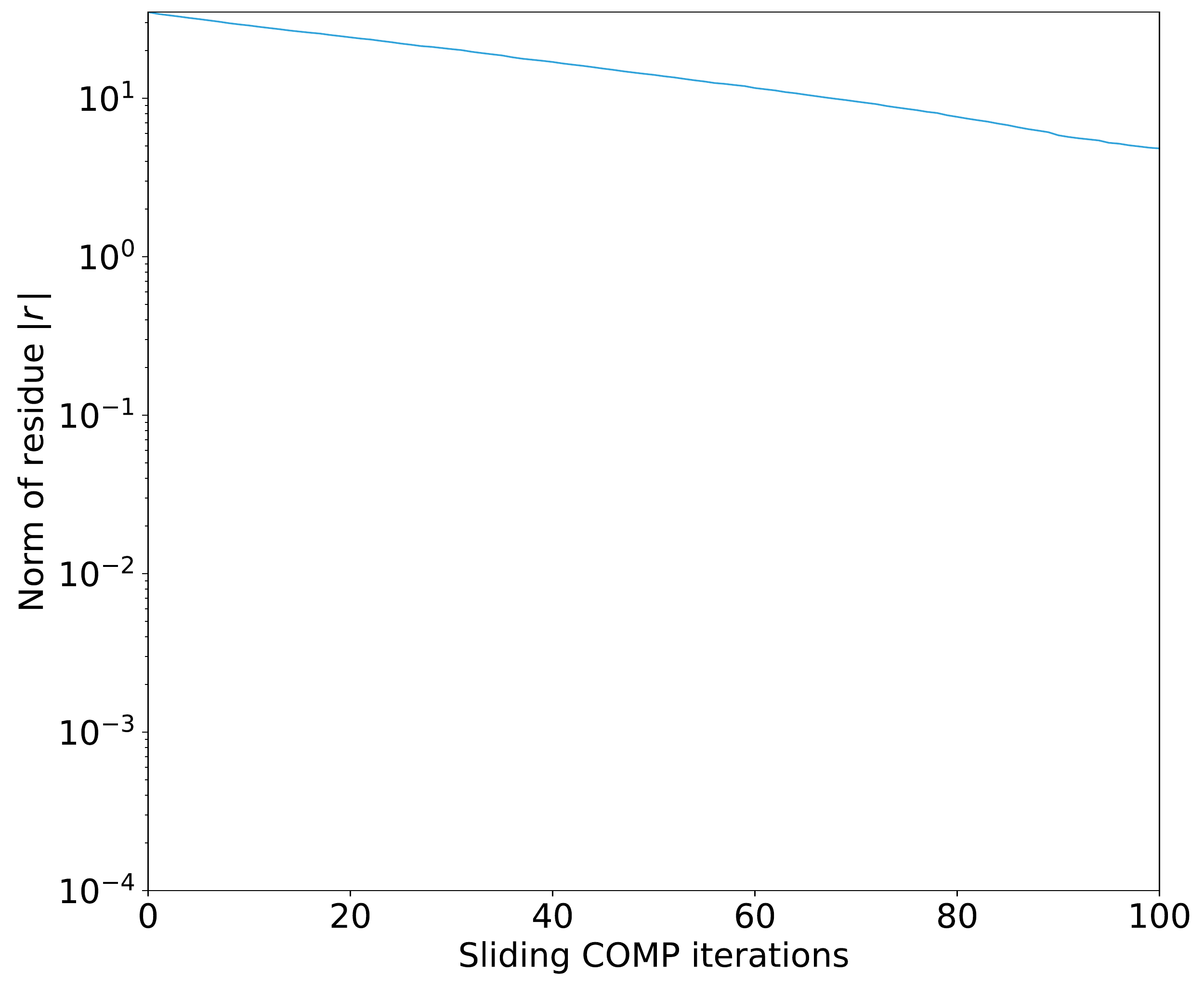}
			\label{fig:2D_COMP_error}
		}
		\subfigure[]{
			\includegraphics[width=0.45\columnwidth]{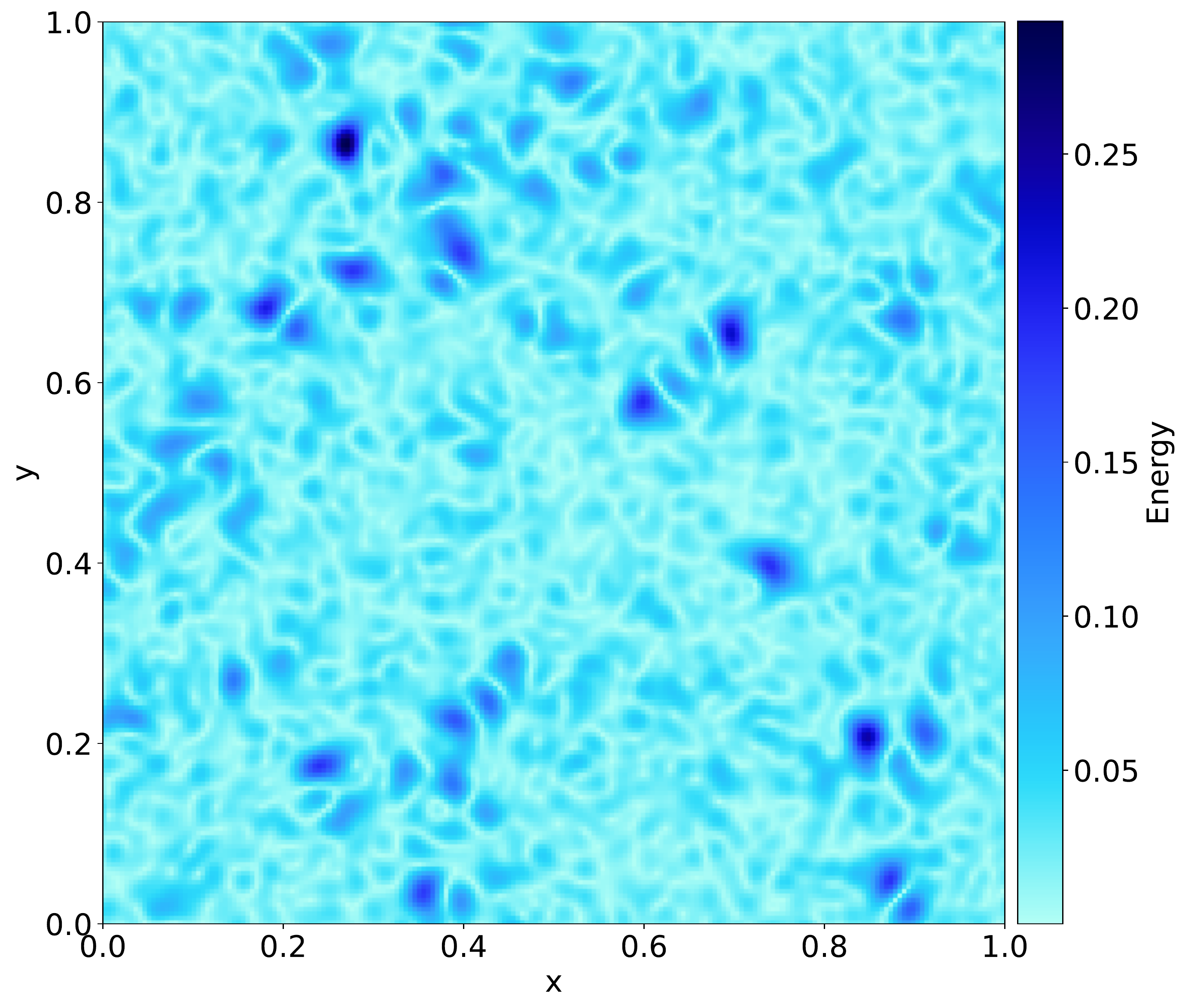}
			\label{fig:2D_COMP_residue_energy}
		}
		\vspace{-.5\baselineskip}
		\caption{
			(a) Ground truth and its back-projection, 
			(b) Ground truth and the estimated signal by Sliding COMP, 
			(c) Norm of the residue after adding each spike by Sliding COMP, 
			(d) Back-projection of the residue by Sliding COMP}
		\label{fig:foobar}
	\end{figure}
	
	In Fig. \ref{fig:2D_gd_energy}, we represent the 2D signal as well as its back-projection. In this figure and the following, the intensity of the colors represent the amplitudes of the spikes. Some spikes close from each other form clusters that are not separated in the back-projection. For some spikes with very low amplitude (close to $ 1 $ in this case), they are barely visible on the back-projection. This means that spikes with low amplitude are more complex to detect using initialization by back-projection.
	
	\subsubsection{Sliding COMP}
	
	In Fig. \ref{fig:2D_COMP_gd_esti},  Sliding COMP recovered almost all spikes with its correct amplitude. Due to the non-convex nature of the first step of COMP, some estimated spikes are stuck in local basins of attraction.
	The computation time of  Sliding COMP is \emph{1h 30min}.

	To compare this result to others, we use the error $ e = \| y - A x_{\text{esti}} \|_{2} $. We call $ r:= y - A x_{\text{esti}} $  the residue. 
	
	As we initialize the estimated signal spike by spike, the norm of the residue $ r $ decreases steadily to attain $ \|r^{(k)}\|_2 \approx 5 $. As we see in Fig. \ref{fig:2D_COMP_error}, the norm of the residue is still decreasing at the end, meaning that the optimization and/or the adding process  has not completely converged to the true solution.  However, if we compare the back-projection of the observed signal $ y $ in Fig. \ref{fig:2D_gd_energy} with the back-projection of the residue $ r^{(k)} $ as in Fig. \ref{fig:2D_COMP_residue_energy}, the order of magnitude of the energy has decreased. This means that Sliding COMP recovered the majority of the signal.
	
	\subsubsection{Over-parametrized COMP}
	
\begin{figure}[htbp]
		\centering
		\subfigure[]{
			\includegraphics[width=0.45\columnwidth]{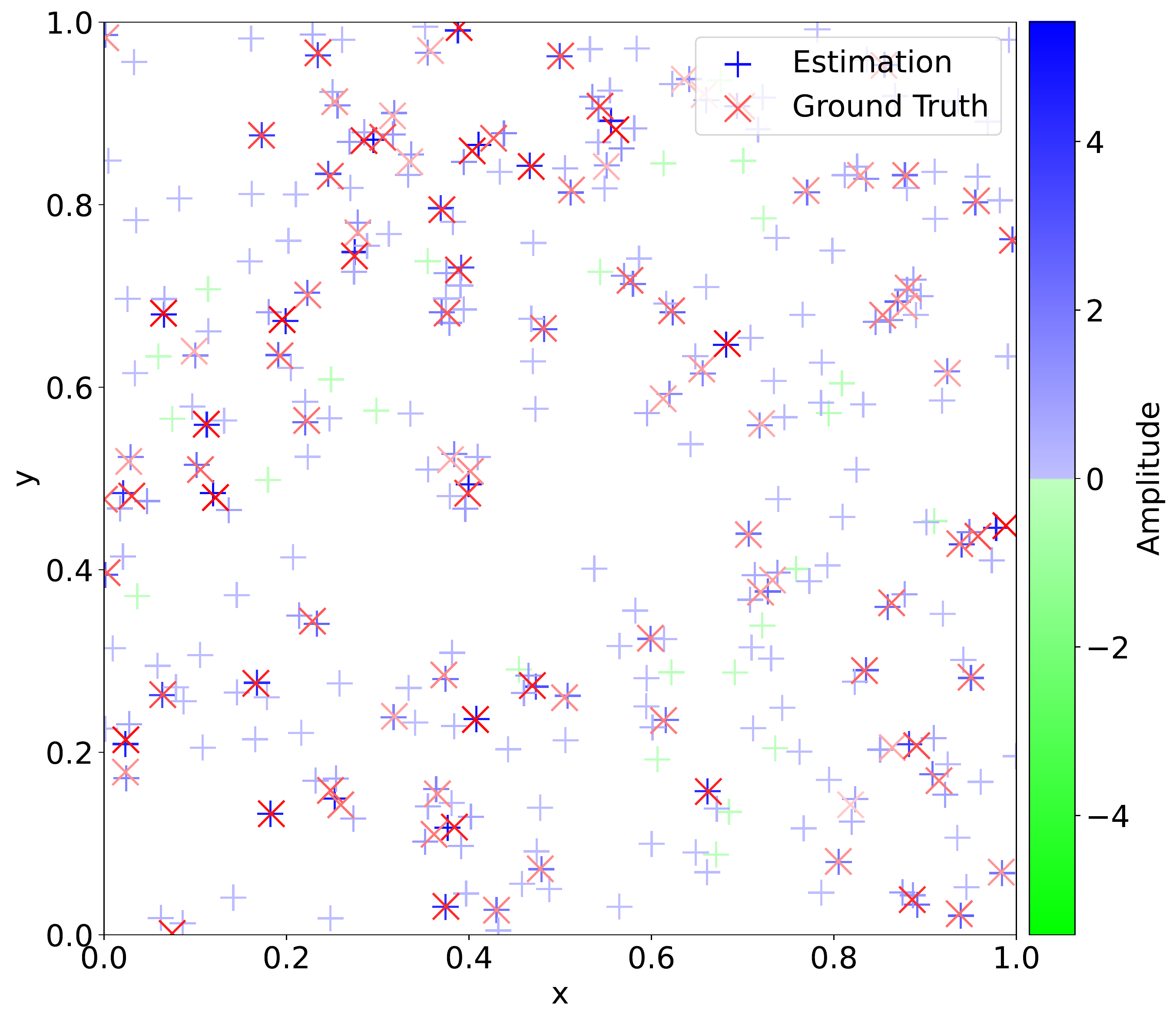}
			\label{fig:2D_SCOMP_gd_init}
		} 
		\subfigure[]{
			\includegraphics[width=0.45\columnwidth]{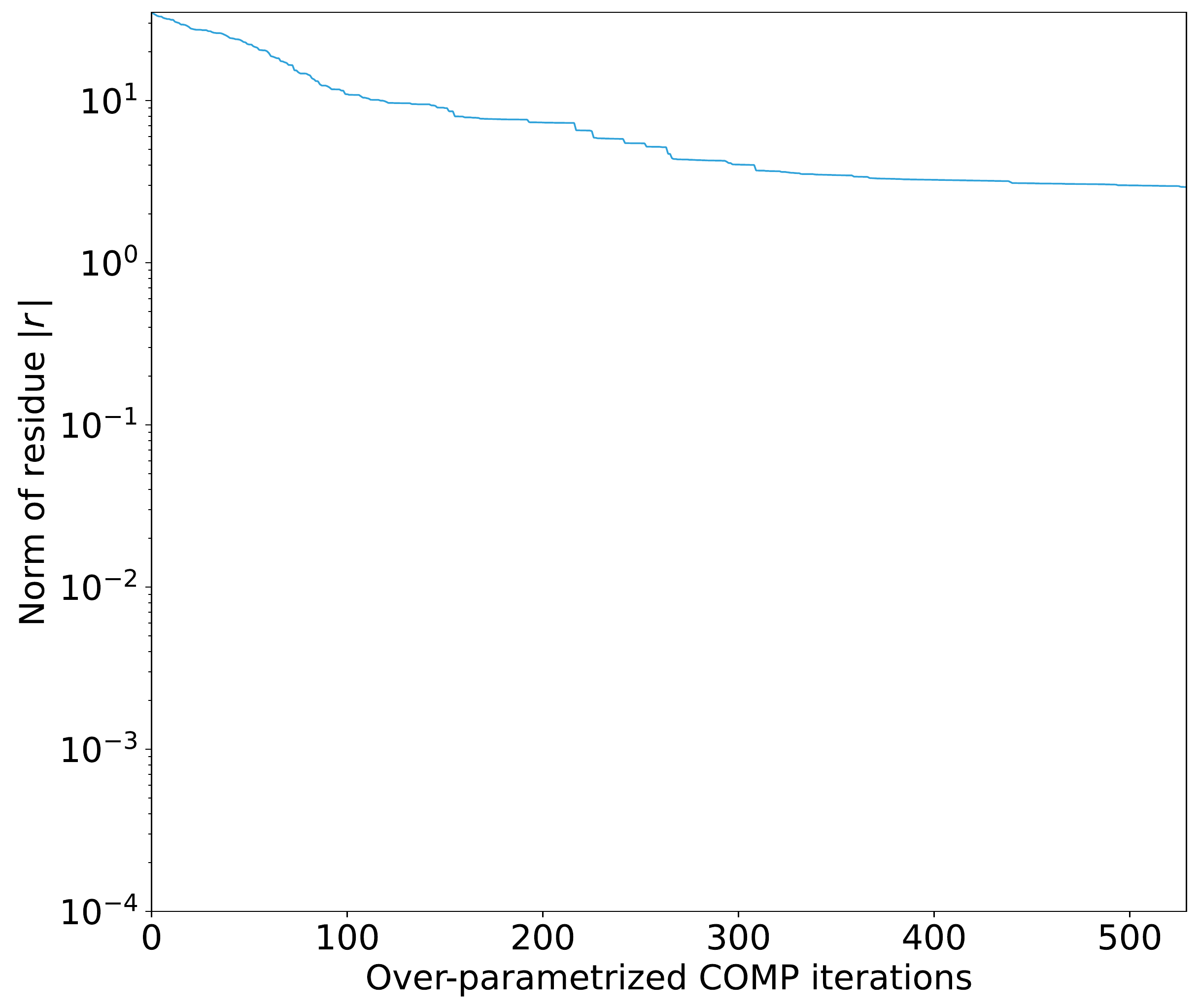}
			\label{fig:2D_SCOMP_error_init}
		} \\[-2ex]
		\subfigure[]{
			\includegraphics[width=0.45\columnwidth]{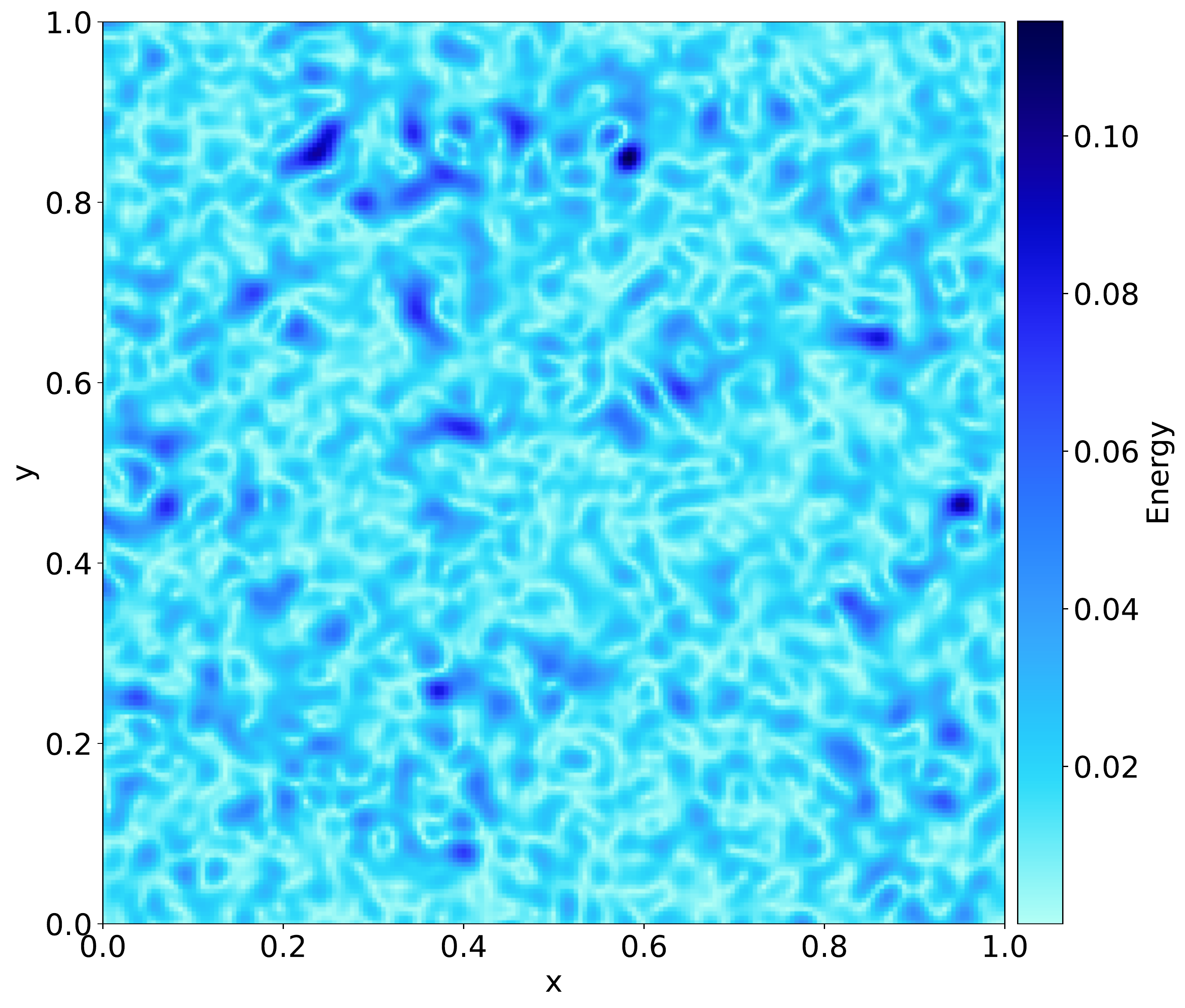}
			\label{fig:2D_SCOMP_residue_init_energy}
		}
		\subfigure[]{
			\includegraphics[width=0.43\columnwidth]{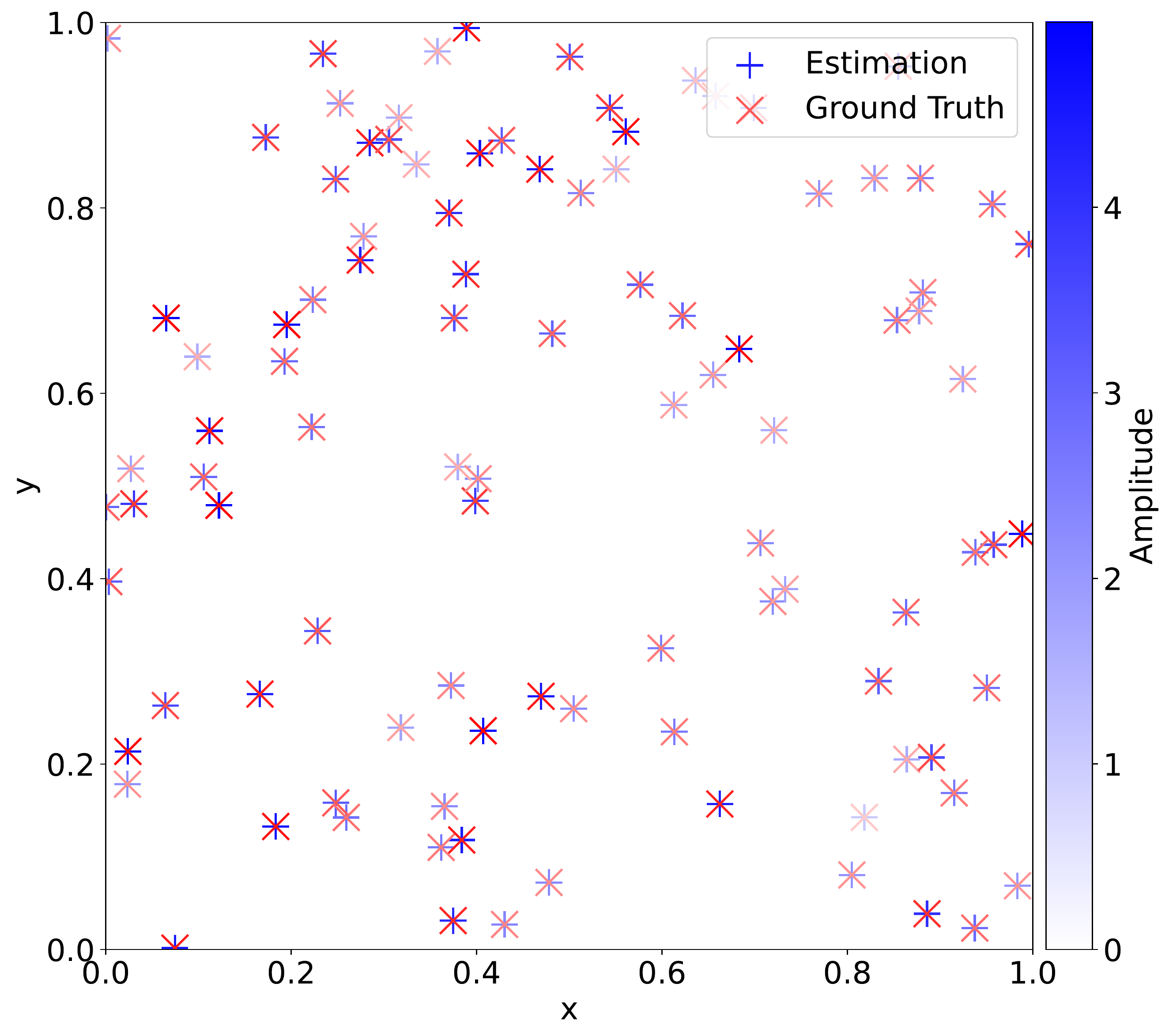}
			\label{fig:2D_SCOMP_gd_esti}
		}
		\vspace{-.5\baselineskip}
		\caption{
			(a) Ground truth and initialized signal by over-parametrized COMP, 
			(b) Norm of the residue after adding each spike by over-parametrized COMP, 
			(c) Back-projection of the initialized residue by over-parametrized COMP, 
			(d) Ground truth and estimated signal by over-parametrized COMP}
		\label{fig:foobar2}
	\end{figure}

	To understand the role of the initialization with the over-parametrized COMP, we show its results without the PGD step in Fig. \ref{fig:2D_SCOMP_gd_init}.
	 We observe that every spikes of the true signal has at least an initialized spike close to it. In some places, we notice a cluster of spikes. This is for these cases that the projection part is needed. The time to compute the initialization is \emph{1min 54s}. We note the low cost of this initialization compared to Sliding COMP.
	We see that in Fig. \ref{fig:2D_SCOMP_error_init},  the norm of the residue is steadily decreasing and that in the last few steps, it stays constant. This is the limit that this method attains and  adding more spikes does not increase the accuracy of the estimation.
	
	By comparing the back-projection of the residue after initialization with the back-projection of the observation, we see in Fig. \ref{fig:2D_SCOMP_residue_init_energy} that the order of magnitude of the error has decreased. Yet there are still spots with a lot of energy meaning that the estimation is not accurate enough. This is why we still need the PGD to optimize the initialized signal.
	
	\subsubsection{PGD after over-parametrized COMP}
	After the optimization step with the PGD, we get the following result in Fig. \ref{fig:2D_SCOMP_gd_esti}.
	All the spikes from the ground truth have been estimated. This procedure took approximately \emph{24min} and it converged after $ 2656 $ iterations of the PGD. We consider that an estimated signal has converged to the ground truth if the norm of the residue is lower that a threshold. With a much smaller calculation time our method was able to recover more accurately the spikes. Improvement in calculation times would be even greater when comparing to the more accurate but slower Sliding COMP with replacement.
		
	\begin{figure}[htbp]
		\centering
		\subfigure[]{
			\includegraphics[width=0.45\columnwidth]{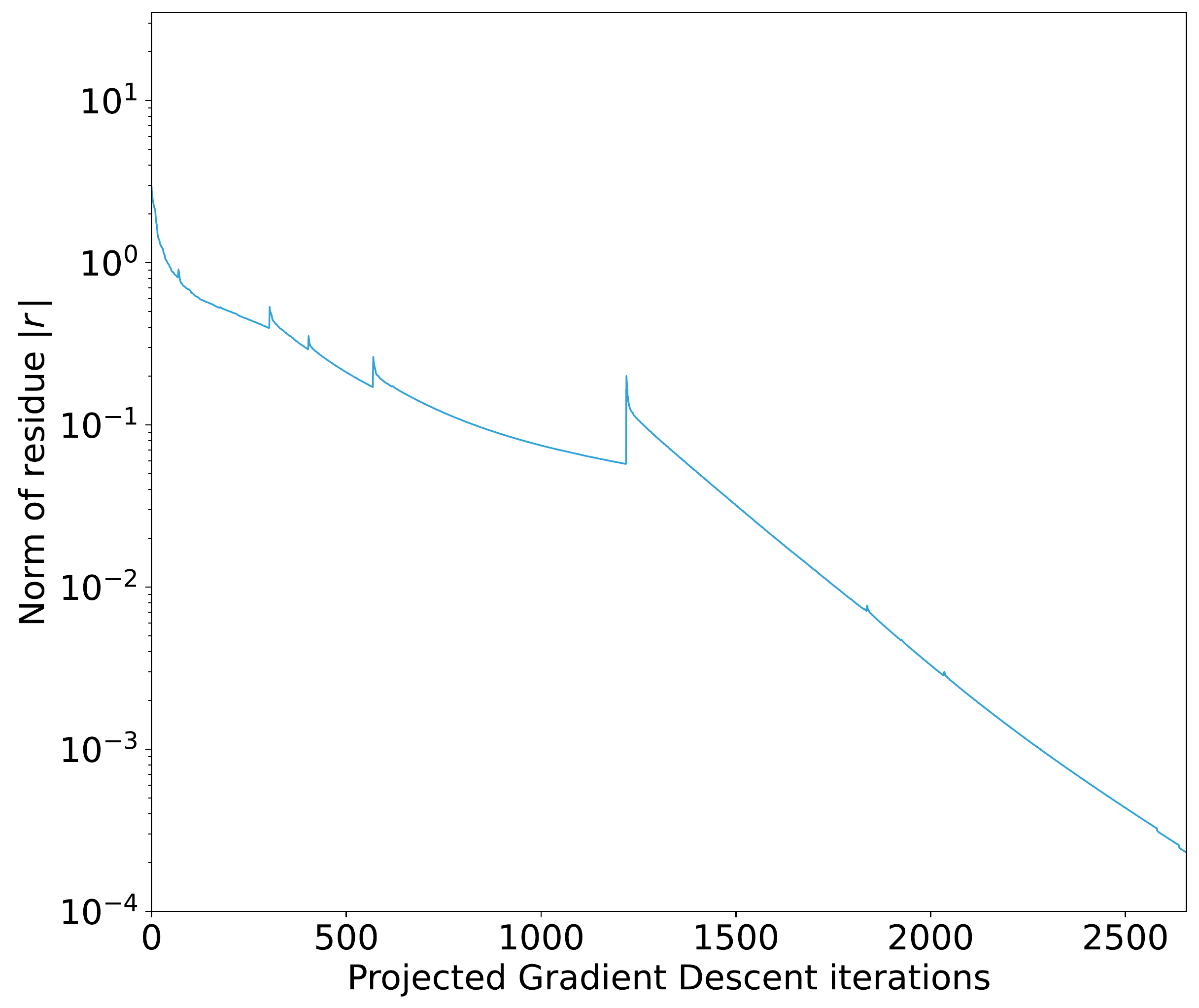}
			\label{fig:2D_SCOMP_error_esti}
		}
		\subfigure[]{
			\includegraphics[width=0.45\columnwidth]{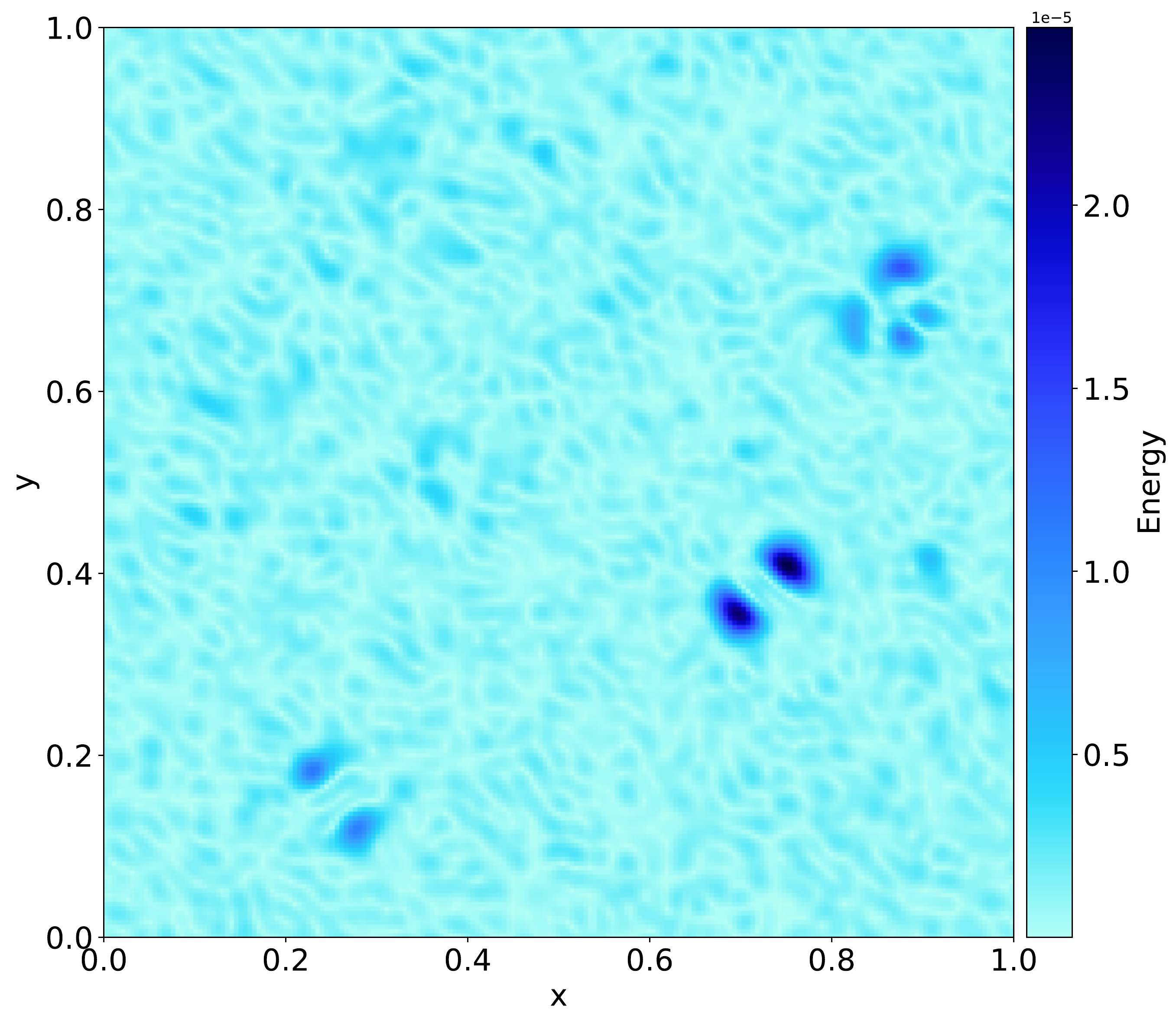}
			\label{fig:2D_SCOMP_residue_esti_energy}
		}
		\vspace{-.5\baselineskip}
		\caption{
			(a) Norm of the residue during the PGD,
			(b) Back-projection of the estimated residue by PGD}
		\label{fig:foobar3}
	\end{figure}

	By observing Fig. \ref{fig:2D_SCOMP_error_esti}, we also deduce that all the signal has been estimated. We note that the norm of the residue may increase sometimes during the PGD. This is a typical phenomenon that must be controlled to be able to prove convergence. Finally, the norm of the residue is close to zero and the back-projection of the residue has a very low energy compared to the original back-projection of the observation of the true signal in Fig. \ref{fig:2D_SCOMP_residue_esti_energy}. We can note the difference of scale with the back-projection of the residue obtained with Sliding COMP.
	
	\subsubsection{Limits of over-parametrization}
	
\begin{figure}[htbp]
		\centering
		\subfigure[]{
			\includegraphics[width=0.4\columnwidth]{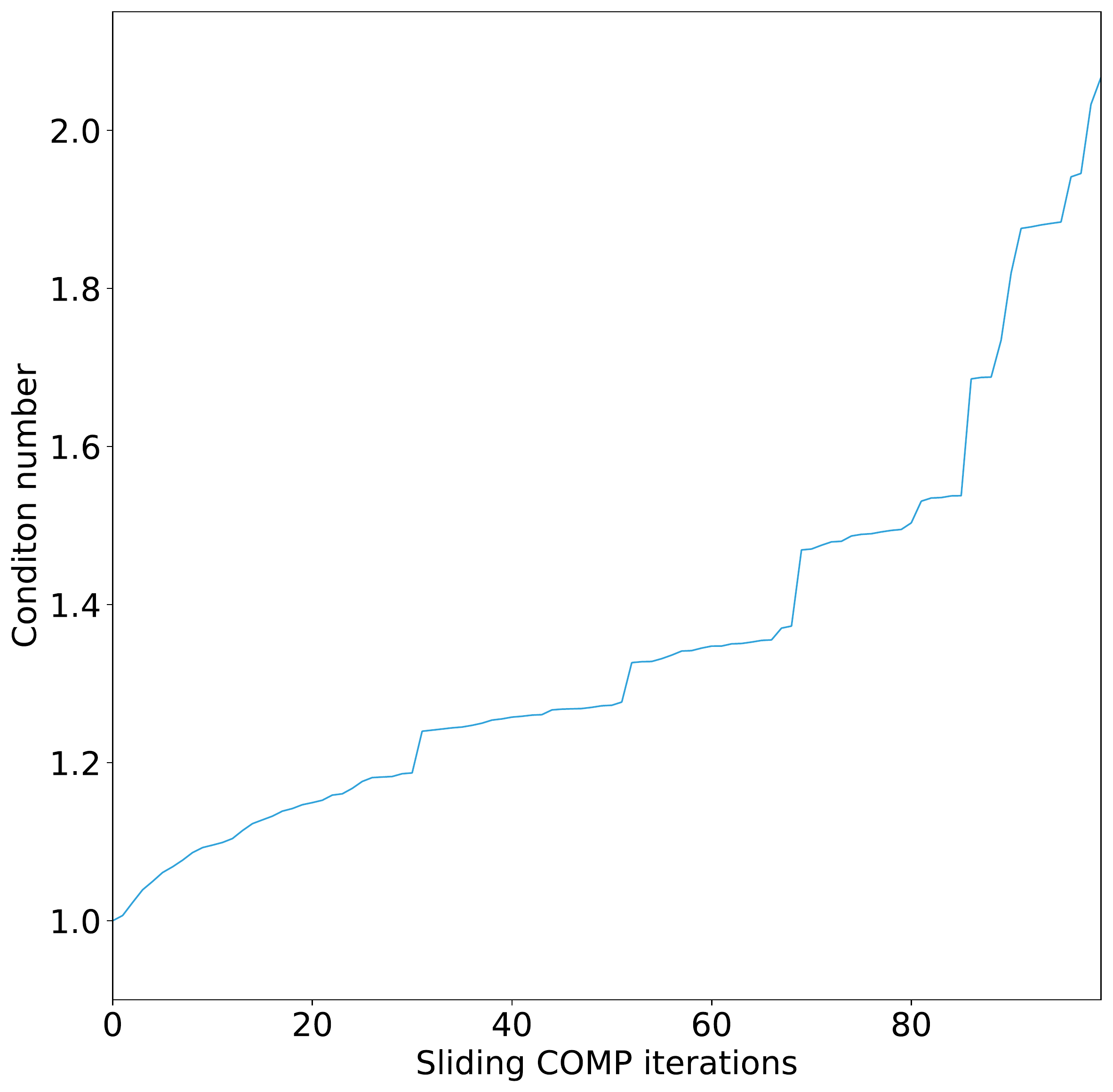}
			\label{fig:2D_COMP_condition_number}
		}
		\subfigure[]{
			\includegraphics[width=0.4\columnwidth]{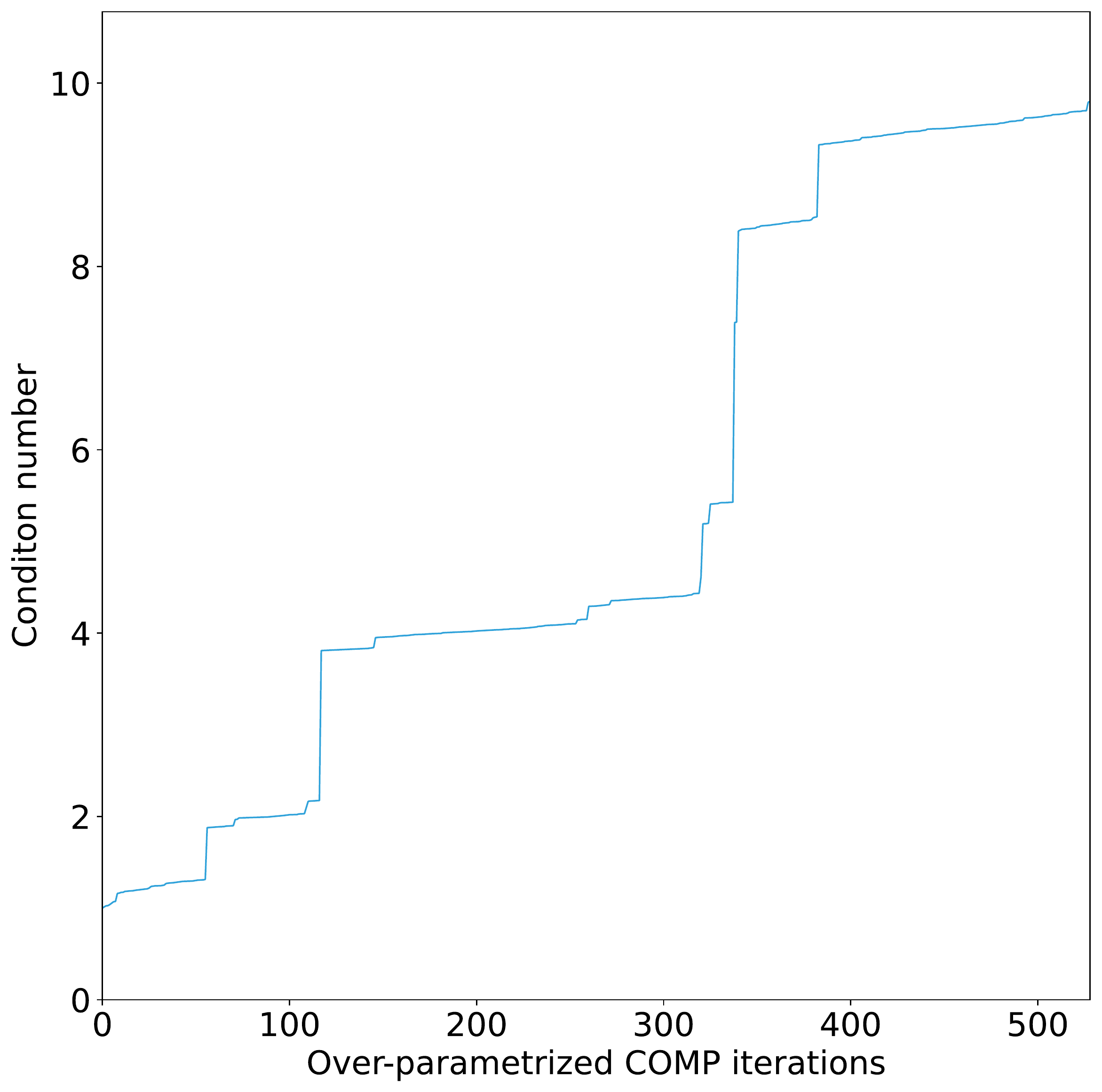}
			\label{fig:2D_SCOMP_condition_number}
		}
		\vspace{-.5\baselineskip}
		\caption{
			(a) Condition number of $ M $ by Sliding COMP, 
			(b) Condition number of $ M $ by over-parametrized COMP}
		\label{fig:fooba}
	\end{figure}

	Computing the amplitudes at each step is done by solving a least-squares $ M $ of size $m \times k$. The more spikes we add, the bigger $ M $ gets. We show that in Figs. \ref{fig:2D_COMP_condition_number} and \ref{fig:2D_SCOMP_condition_number}, the condition number of $ M $ at the end of the over-parametrized COMP is higher than for  Sliding COMP. This means that too much over-parametrization leads to a condition number so high that it can induce errors in the computation. This also shows that the amount of over-parametrization by COMP cannot be arbitrarily large.
	\vspace{-8pt}
	\subsection{Comparisons on the 3D case}
\begin{figure}[t]
		\centering
		\subfigure[]{
			\includegraphics[width=0.45\columnwidth]{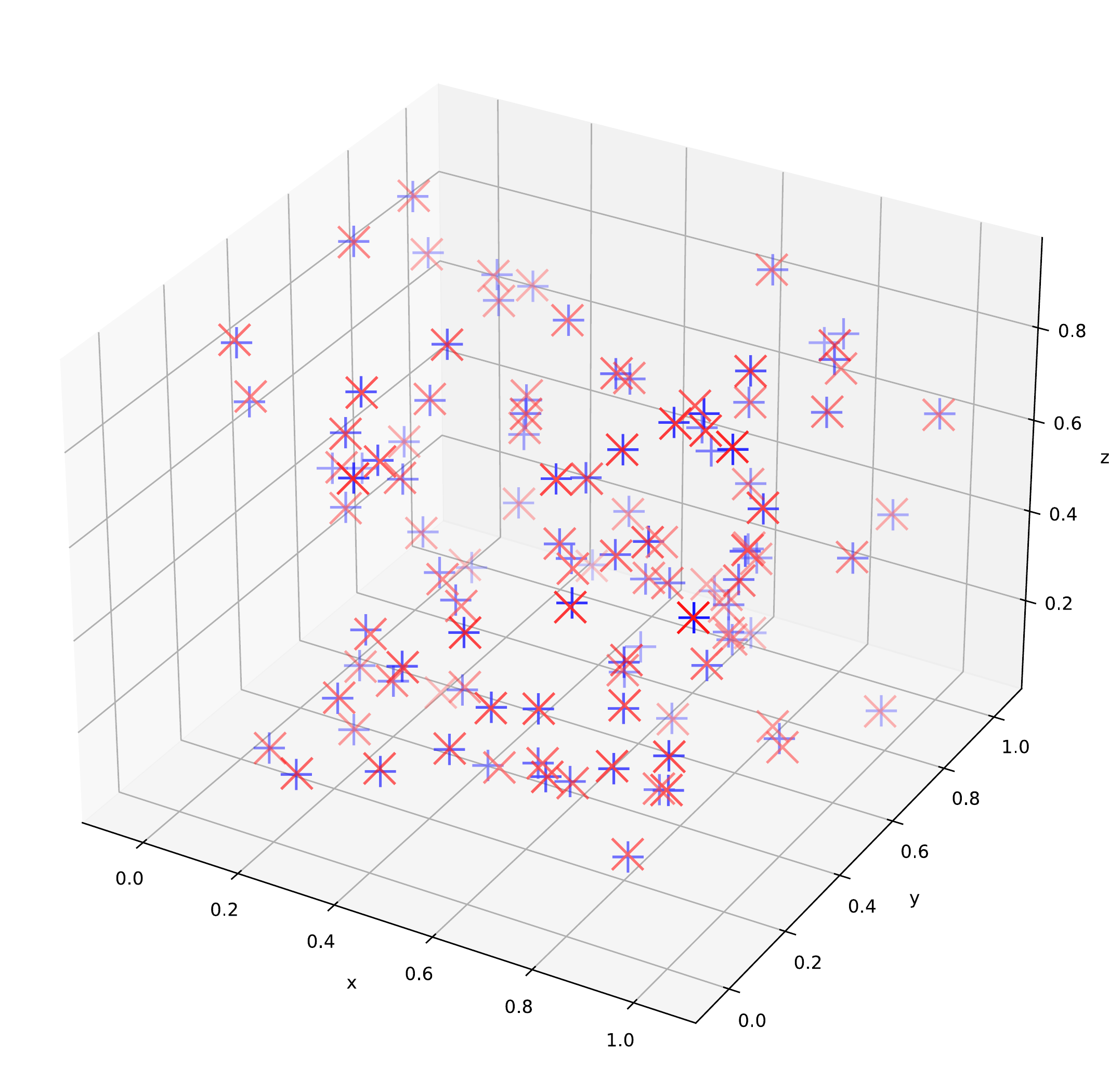}
			\label{fig:3D_gd_esti_COMP}
		}
		\subfigure[]{
			\includegraphics[width=0.45\columnwidth]{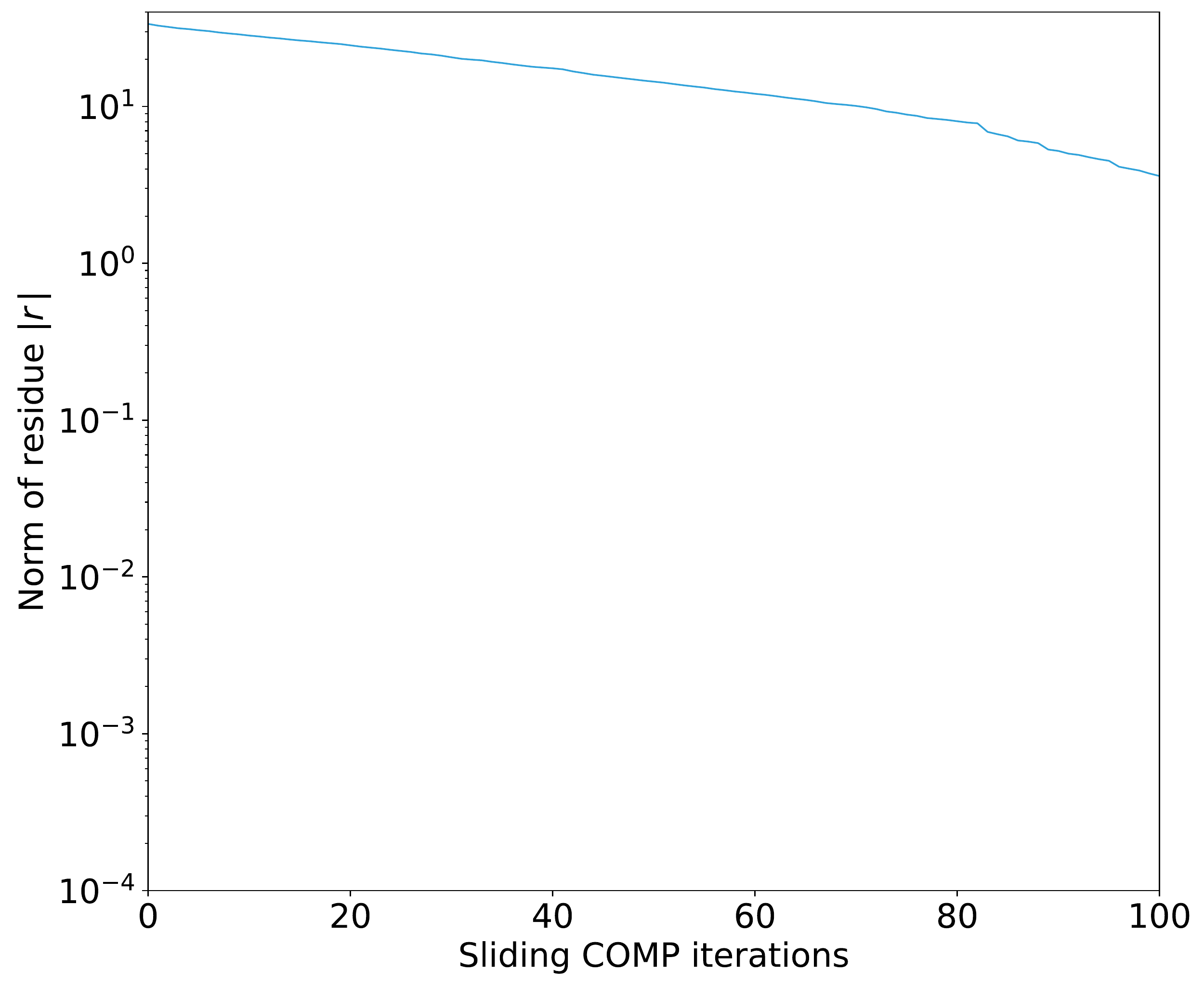}
			\label{fig:3D_COMP_error}
		}
		\vspace{-.5\baselineskip}
		\caption{
			(a) Ground truth and estimation by Sliding COMP,
			(b) Norm of the residue during the Sliding COMP}
		\label{fig:foobar4}
	\end{figure}

	For the  the recovery of a signal in 3D (represented in red in Fig. \ref{fig:3D_gd_esti_COMP}), we perform the same  analysis. The same parameters used in the 2D case are applied to some exceptions which are the following. The minimum separation between spikes is set to $ \epsilon_{\text{dist}} = 0.05 $.  Furthermore, since the measurements depend in our case on $ \epsilon_{\text{dist}} $, they follow a normal distribution $ \mathcal{N}(0, c^{2}) $ with $ c = \frac{1}{0.05} \approx \frac{1}{\epsilon_{\text{dist}}} $. 
	
	\begin{figure}[]
		\centering
		\subfigure[]{
			\includegraphics[width=0.45\columnwidth]{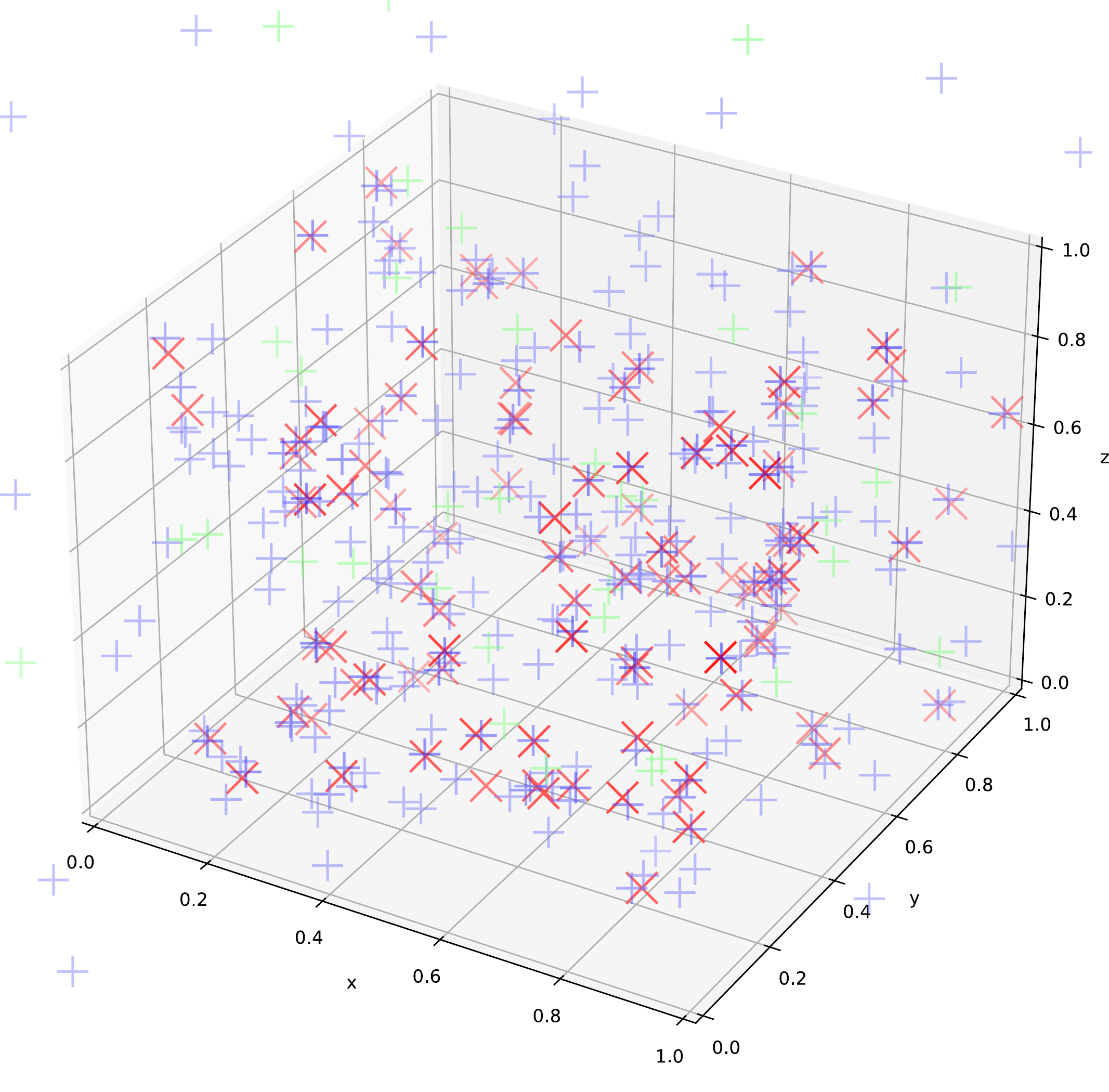}
			\label{fig:3D_SCOMP_gd_init}
		}
		\subfigure[]{
			\includegraphics[width=0.45\columnwidth]{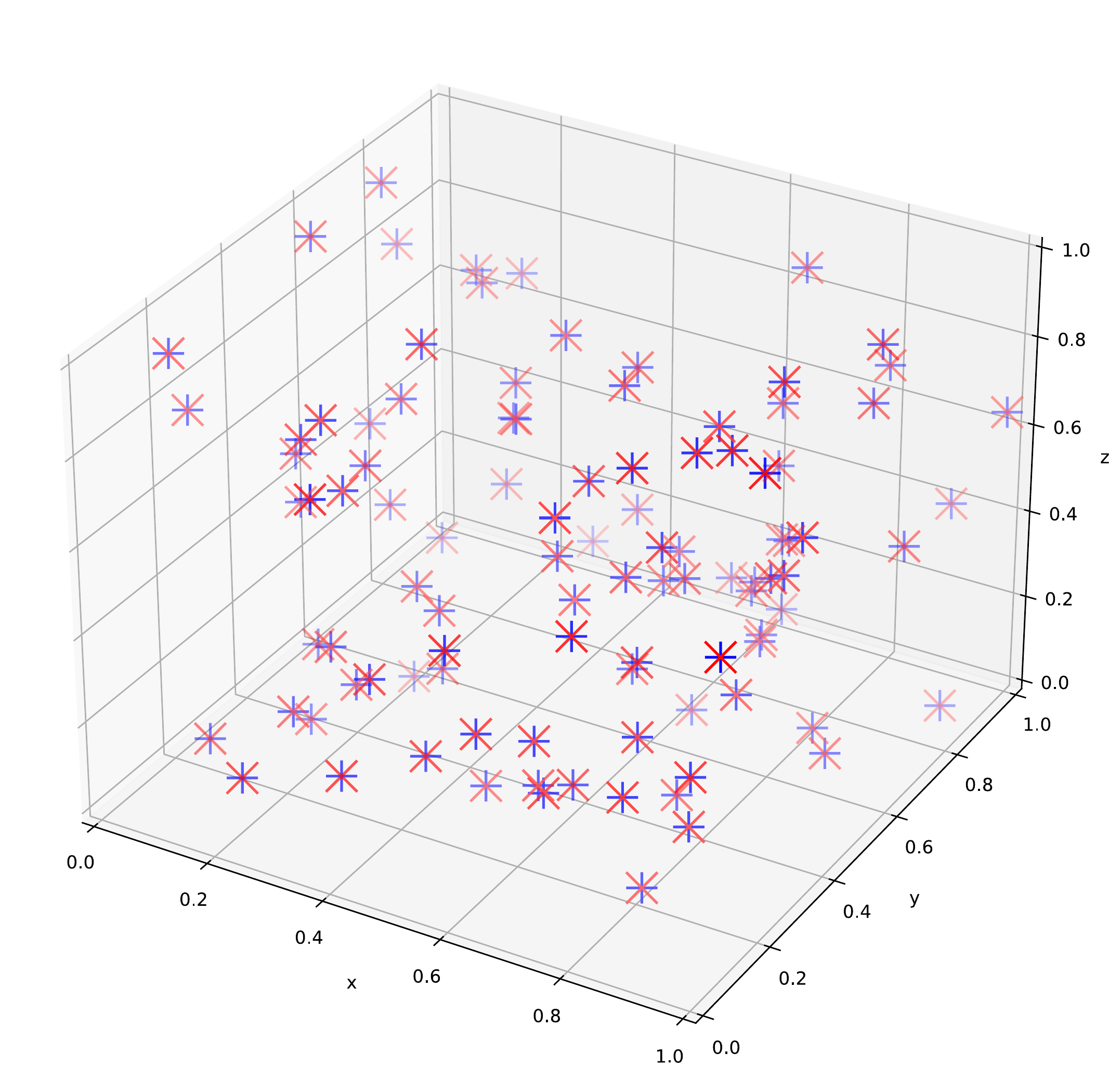}
			\label{fig:3D_SCOMP_gd_esti}
		} \\[-2ex]
		\subfigure[]{
			\includegraphics[width=0.45\columnwidth]{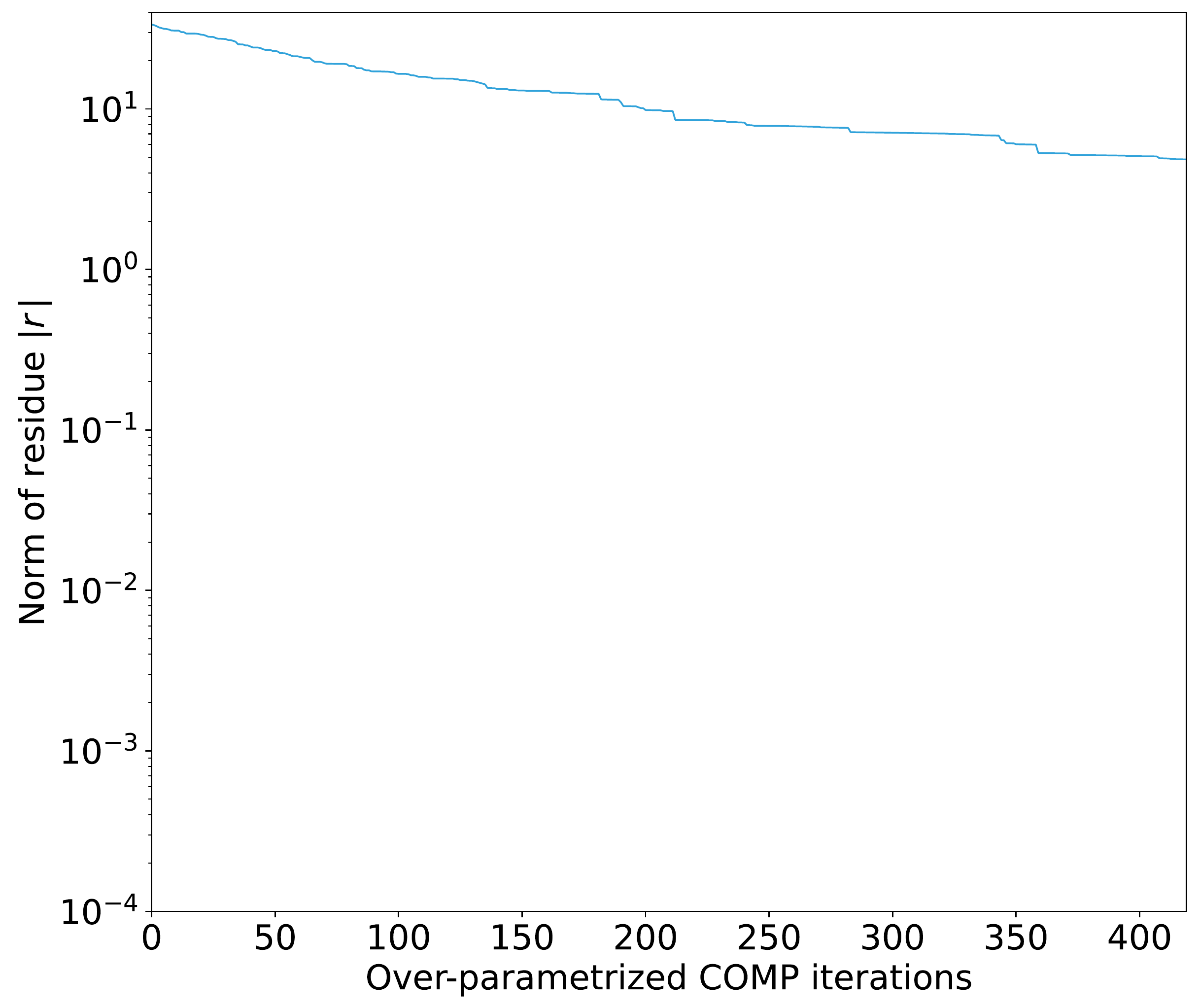}
			\label{fig:3D_SCOMP_error_init}
		}
		\subfigure[]{
		\includegraphics[width=0.45\columnwidth]{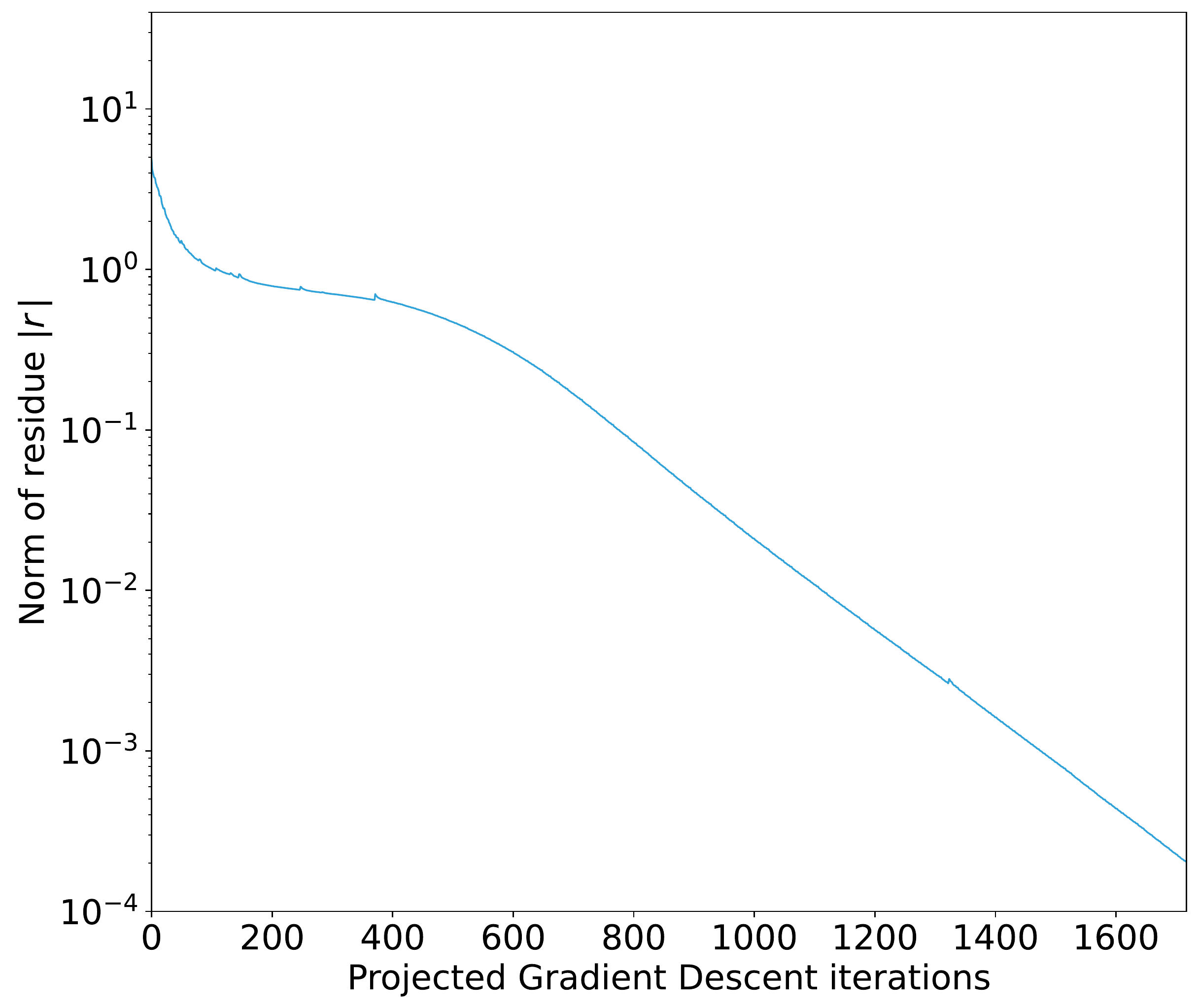}
		\label{fig:3D_SCOMP_error_esti}
		}
		\vspace{-.5\baselineskip}
		\caption{
			(a) Ground truth and initialization by over-parametrized COMP,
			(b) Ground truth and estimation by PGD,
			(c) Norm of the residue during the over-parametrized COMP,
			(d) Norm of the residue during the PGD}
		\label{fig:foobar5}
	\end{figure}

	\vspace{3pt}
	\subsubsection{Sliding COMP}
	We see that in Fig. \ref{fig:3D_gd_esti_COMP}, most spikes are also recovered with some exceptions. It is important to note that in some place, two or more spikes from the ground truth signal have been estimated by a single spike of the estimated signal. The amplitude of this spike is roughly the sum of the amplitudes of the spikes in the cluster. Sliding COMP computed this estimation in approximately \emph{1h 30min}. For the norm of the residue in Fig. \ref{fig:3D_COMP_error}, it is decreasing over the iterations but does not attain $ 0 $ after adding $ 100 $ spikes. We also note that this time is not too far apart from the time in the 2D case. We can deduce that this family of algorithm is not too dependent on the dimension of the space. On the contrary to methods needing a grid like in \cite{traonmilin:projected_gradient_descent}.
	
	\vspace{3pt}
	\subsubsection{Over-parametrized COMP and PGD}
	After the initialization with over-parametrized COMP, we get the signal in Fig. \ref{fig:3D_SCOMP_gd_init}. We note that all the spikes from the ground truth have at least one spike from the initialized signal. Moreover, some spikes have a negative amplitude (in green). 
	This step took approximately \emph{1min} after adding $ 418 $ spikes. In Fig. \ref{fig:3D_SCOMP_error_init}, we note that the norm of the residue decreases until it reaches a threshold. Adding even more spikes does not increase the accuracy of the estimation. Same as Sliding COMP, this algorithm seems not to be too dependent from the number of dimensions.
	
	For the projected gradient descent, we note that after $ 1716 $ iterations, the energy of the system reached a threshold close to $ 0 $ to consider convergence in Fig. \ref{fig:3D_SCOMP_error_esti}. It took approximately \emph{13min 45s}. As the cost of the iteration is similar in 2D and 3D, the fewer number of iterations needed to converge in 3D could be explained by the greater distance between spikes. We get as a final result the signal in Fig. \ref{fig:3D_SCOMP_gd_esti}.	
	
	\section{Discussion/Conclusion}
	
	We showed that Projected Gradient Descent initialized by over-parametrized COMP without sliding  leads to better results in faster times than Sliding COMP. Indeed, it provides a way to catch all spikes from the ground truth and without needing to know precisely the number of true spikes to recover. We have shown the success of our method in both 2D and 3D and expect similar results for examples in large dimensions. 
	To recapitulate, the Table \ref{tab:recap_times} shows the computation times of each algorithms.

	\begin{table}
		\centering
		\caption{Summary of computation times}
		\vspace{-.5\baselineskip}
		\begin{tabular}{c|cc}
			& \multicolumn{2}{c}{\begin{tabular}[c]{@{}c@{}}Time (in min.)\end{tabular}} \\ \hline
			Dimensions                 & \multicolumn{1}{c|}{2D} & 3D \\ \hline
			Sliding COMP               & \multicolumn{1}{c|}{90} & 90 \\ \hline
			Over-parametrized COMP     & \multicolumn{1}{c|}{2}  & 1  \\ \hline
			Projected Gradient Descent & \multicolumn{1}{c|}{24} & 14 \\ \hline
		\end{tabular}
	
		\label{tab:recap_times}
	\end{table}

	Although these are some promising experimental results, there are still no theoretical guarantees for this new method. However, some very strong quantitative and qualitative insights already exist \cite{traonmilin:hal-01938239,traonmilin:2009.08670}, and hope for the possible extension of proofs from the finite dimensional domain to our case gives interesting potential leads for future work.
	
	\section{Acknowledgments}
	
	This work was supported by the French National Research Agency (ANR) under reference ANR-20-CE40-0001 (EFFIREG project).
	
	\bibliographystyle{IEEEtran}
	\bibliography{bibliography.bib}

\end{document}